\documentclass[10pt]{article}
\usepackage{amsmath,amssymb,cite,color,epsfig,graphicx,fullpage,hyperref}
\hypersetup{colorlinks=true,linkcolor=blue,citecolor=blue,urlcolor=blue}
\newcommand{\bigO}{{\mathcal O}}

\title{Different ways of looking at the force between two nano crystals}
\author{Alexander Lange$^{12}$, Fabian Danecker$^1$, Gernot Bauer$^1$, Nadezda Gribova$^{13}$, Joachim Gross$^1$\\[2ex]
$^{1)}$\,Institute of Thermodynamics and Thermal Process Engineering\\
$^{2)}$\,Institute of Chemical Process Engineering\\
$^{3)}$\,Institute of Computational Physics\\[2ex]
University of Stuttgart, Germany\\[2ex]
{E-mail: lange@itt.uni-stuttgart.de}
}

\begin{document}

\maketitle

\begin{abstract}

The potential of mean force (PMF) between two nano crystals (NCs) represents an effective interaction potential that can be used to study the assembly of NCs to various superstructures. For a given temperature, the effective interaction is obtained best from molecular dynamics simulations.
Based on a density functional approach, this study proposes three methods of predicting the PMF for any temperature based on a single molecular dynamics simulation for one temperature. The three methods construct the PMF by considering the ligands as an ideal gas, as hard-sphere chains, or as Lennard-Jones interaction sites.
To apply this methodology, the density of the interaction centers must be extracted from the simulation data. For the ideal gas model, a straightforward sampling procedure with a fixed lattice in space leads to free energies that are too large in order to consistently explain the simulation data for different temperatures.
Naive sampling does not account for the small momenta added to the NCs when coupled to a thermostat. A method is proposed that corrects for the unphysical steps during the simulation. The ideal gas contribution computed for the corrected density is significantly smaller than the one obtained from naive sampling and can thus explain the temperature dependence of the PMF correctly.
For the hard-sphere chain model, where a weighted density is used, the correction of the particle density is not essential. However, the PMF calculated based on the corrected density confirms our approach.
All three models predict PMF curves in very good agreement with simulation results, but they differ in the number of input parameters and the computational effort. Based on the modeling results, we predict the existence of an additional attractive force at small distances of the NCs --- a depletion force.

\paragraph*{PACS} 34.20.Gj; 65.80.-g; 82.60.Qr

\paragraph*{Keywords} capped nano crystals; thiol ligands; gold core; potential of mean force; perturbation theory; depletion attraction

\end{abstract}

\section{Introduction}

Nano crystals (NCs) --- particles the size of a few nanometers, consisting of many thousand atoms, usually capped with ligands --- are building blocks of new materials with applications in various fields of modern engineering \cite{KCR08}. Much about them, especially about their interaction with other NCs, is known from experiments \cite{WGASP} and molecular simulations \cite{SPV08}, including simplified models \cite{LYEB11}. Due to their complex structure and interaction behavior, a good theoretical knowledge is often missing. 
The precise shape of pair- and many-body-potentials, which determine the assembly of NCs into superstructures, is not well understood theoretically and is so far obtained best by molecular simulations \cite{KPSC09,SV09}.

In this article we investigate the effective pair potential employing models of density functional theory (DFT) while relying on data from molecular dynamics (MD) simulations.
The procedure allows for an accurate prediction of effective pair potentials at temperatures other than the temperature considered for the MD simulation. 
To assess our approach,
we run simulations with two thiol capped gold NCs for various temperatures (Fig.~\ref{fig1}).
Due to the large number of interaction sites (1--10k)
and the requirement of sufficiently long simulation runs
computations of this kind are expensive. Therefore, we try to reduce the number of necessary simulations and use all the information that is generated --- not only the final estimates of the mean force.

For instance, we analyze trajectory data from our MD simulations, determine the density of the relevant interaction sites
at the ligand caps and apply their density to various fluid models of density functional theory (DFT).
In doing so we neglect some of the NCs' structure. For example, considering a chain model, we neglect the adsorption of the thiol ligand to the gold core, and considering an ideal gas, we even neglect molecular forces. How can such an approach be justified?

The potential of mean force (PMF) will be represented by the difference of two Helmholtz free energies \cite{J97}; the minuend is the free energy of the two NCs  at focal distance, the subtrahend is the free energy of the NCs at infinite distance.
Therefore, any interaction that does not vary in strength when changing the NCs' distance to each other will not contribute to the PMF. 
One example are the forces by which the ligands are absorbed to the gold surface, another example are the chemical bonds between the atoms in the thiols.
That is, when calculating the PMF, one can neglect most of the NCs' structure.

\section{Methods}

\subsection{MD simulations}

\begin{figure}
\center
\hspace{-11em}\\[-5ex]
\parbox[t]{7cm}{{\Large\sf A}\\[-4ex]
\hspace*{1.5em}
\includegraphics[height=32ex]{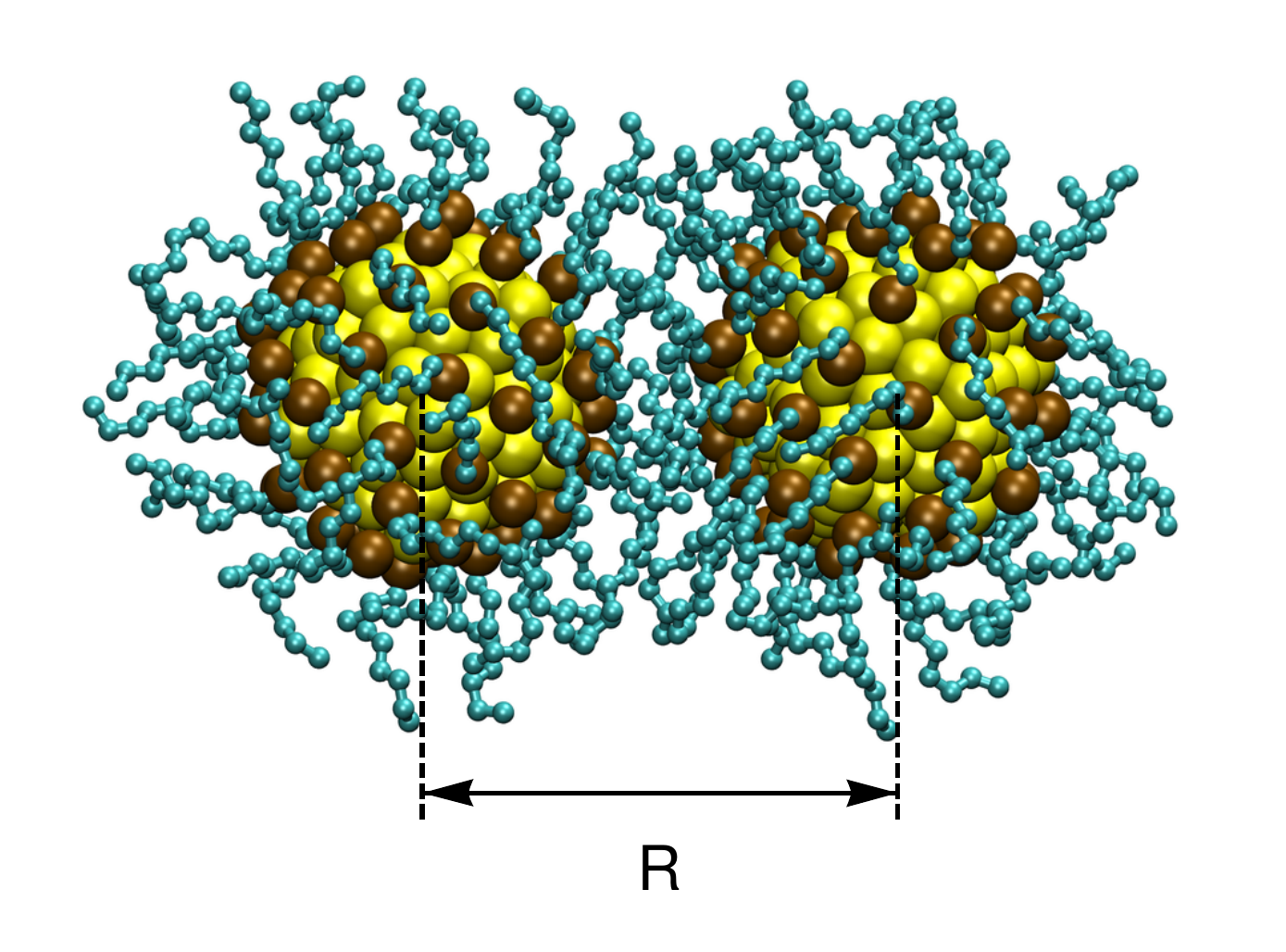}}
\hspace*{2.5em}
\parbox[t]{7cm}{{\Large\sf B}\\[-2ex]
\hspace*{1.em}
\includegraphics[height=29ex]{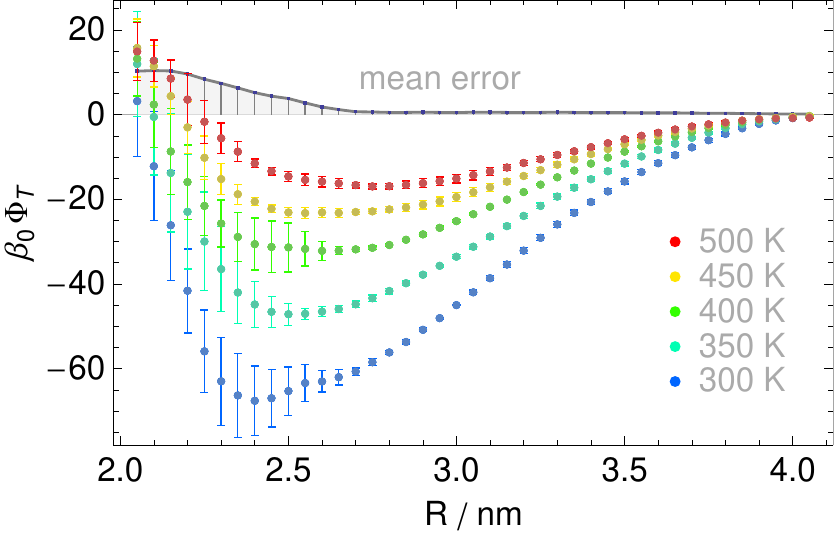}
}\hspace*{1.5em}
\caption{Two exemplar NCs and their PMF.
Panel A illustrates the considered system of two 
$\text{Au}_{147}(\text{SC}_8)_{58}$ NCs at distance $R$.
The NCs have a gold core consisting of 147 Au atoms (forming an icosahedra, yellow), 58 alkanethiol ligands consisting of one $\text{SH}$- (brown) and eight $\text{CH}_x$-groups (cyan).
Panel B shows the PMFs for five temperatures as obtained from MD simulations as well as the statistical errors.
\label{fig1}}
\end{figure}

For a set of five temperatures $T$, constrained MD simulations have been performed to determine the PMF $\Phi_T(R)$ between two exemplar NCs (Fig.~\ref{fig1}).
Following a united-atom approach with the Lennard-Jones force fields from Schapotschnikow et al.~\cite{SPV08}, we consider the $\text{SH}$-, $\text{CH}_2$- and $\text{CH}_3$-groups in the thiol ligands as single interaction sites.
Simulations were conducted as described in our previous work \cite{BLGHG14}, using the GROMACS software \cite{PPSLB13} with a leap-frog stochastic dynamics integrator of step size $\Delta t=2\,\text{ps}$ and friction constant $\gamma=.5\,\text{ps}^{-1}$.
This integrator also acts as a thermostat.
For every temperature ($T=300\,\text{K},\dots,500\,\text{K}$), 10 runs were performed in the NVT ensemble, each with an equilibration time of 1\,ns and a total runtime of 
$t=10\,\text{ns}$, providing us with trajectories of 5000 coordinate snapshots.

The centers of mass of the two NCs were aligned with the $x$-axis of the simulation box and their distance $r_{12}$ was fixed during a simulation by freezing the gold atoms in space. Only in between simulations the distance was changed, incrementally ($\Delta r_{12}=.5\,\text{nm}$), from large ($r_{12}^{\max}=5\,\text{nm}>d_c+2L$) to small distances ($r_{12}^{\min}=2\,\text{nm}\gtrsim d_c$); $d_c=1.8\,\text{nm}$ is the diameter of the core and $L=1.1\,\text{nm}$ is the length of the ligands.
Here we make two implicit assumptions \cite{BLGHG14}:
(i) the rotation of the NCs is negligible \cite{SPV08} and
(ii) the core-to-core interaction is unimportant compared to the ligand-to-ligand and ligand-to-core interactions \cite{TB05}.

To obtain the PMF, we evaluated the forces, ${\bf F}_1$ and ${\bf F}_2$, that act on the centers of mass of the gold cores over the course of a simulation run. The mean force between the two NCs is given by
\begin{align}
F_m(r)=\left\langle\frac{{\bf F}_1-{\bf F}_2}{2}\cdot{\bf e}_x\right\rangle_{r_{12}=r}\,,
\end{align}
where the angular bracket denotes the average in the constrained canonical ensemble and ${\bf e}_x$ the unit vector in $x$-direction. The potential at distance $r_{12}=R$ results from integration over larger distances,
\begin{align}
\Phi_T(R)=\int_{R}^\infty F_m(r)\,dr\,,
\end{align}
which in our discrete setting translates into $\Phi_T(R)=\sum_{n=0}^{(r_{12}^{\max}-R)/\Delta r_{12}}F_m(R+n\Delta r_{12})\Delta r_{12}$\,, with $F_m(r_{12}^{\max})=0$.
Consequently, statistical errors become larger towards smaller distances $R$ (cf.~Fig.~\ref{fig1}B).

\subsection{Thermodynamics}

As a special case of Jarzynski's non-equilibrium equality \cite{J97}, the PMF obtained from our MD simulations can be represented by the difference of two Helmholtz free energies,
\begin{align}\label{eq:PhiR}
\Phi_T(R)=A(R)-A(\infty)\,,
\end{align}
describing thermodynamic states of the NCs separated by a certain distance $R$ \cite{BLGHG14}.
In the context of the simulations,
the limit $A(\infty)$ in Eq.~\eqref{eq:PhiR} corresponds to a value at finite distance, $A(r_{12}^{\max})$.

Most naturally \cite{Z54}, the Helmholtz energy is expanded in powers of the inverse temperature $\beta=1/(k_BT)$,
\begin{align}\label{eqHT1}
 \beta A=\sum_{n=0}^\infty \beta^n a_n\,,
\end{align}
where $k_B$ defines Boltzmann's constant.
Even if not indicated, the coefficients $a_n$ also depend on $\beta$,
but in the considered temperature range the dependence is unimportant.

\begin{figure}
\center
\parbox[t]{7cm}{{\Large\sf A}\\[-2ex]
\hspace*{.2em}
\includegraphics[height=29ex]{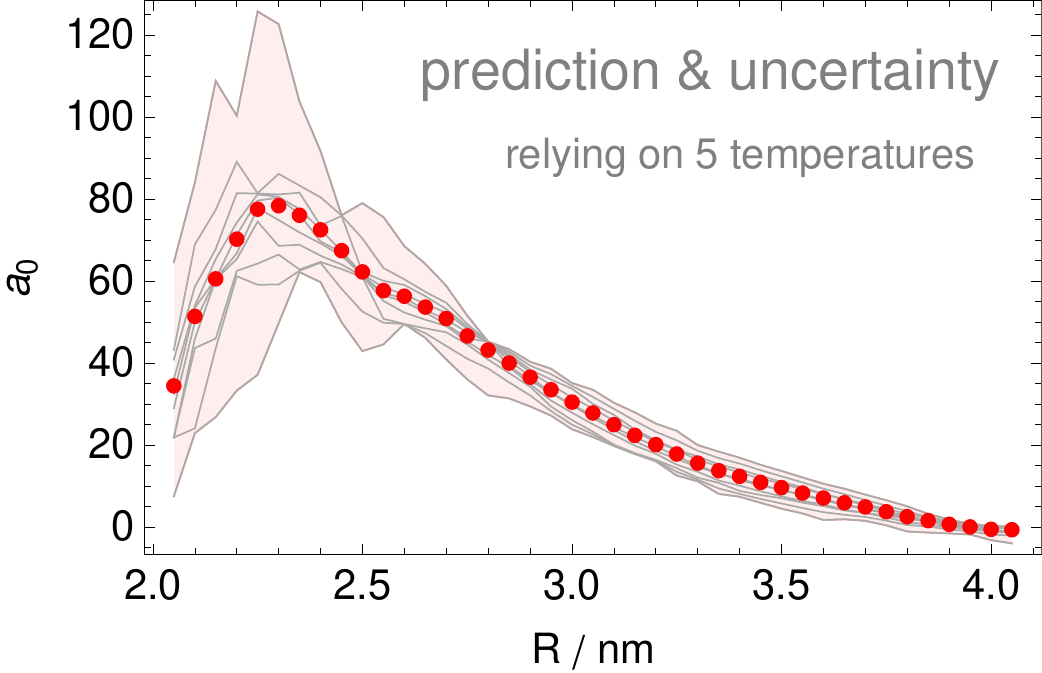}}
\hspace*{1.5em}
\parbox[t]{7cm}{{\Large\sf B}\\[-2ex]
\includegraphics[height=29ex]{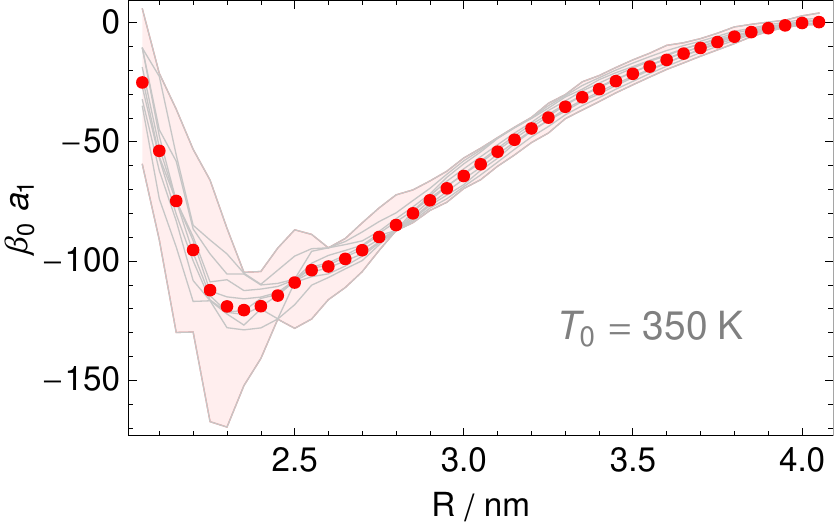}
}\\[2ex]
\parbox[t]{7cm}{{\Large\sf C}\\[-2ex]
\includegraphics[height=30ex]{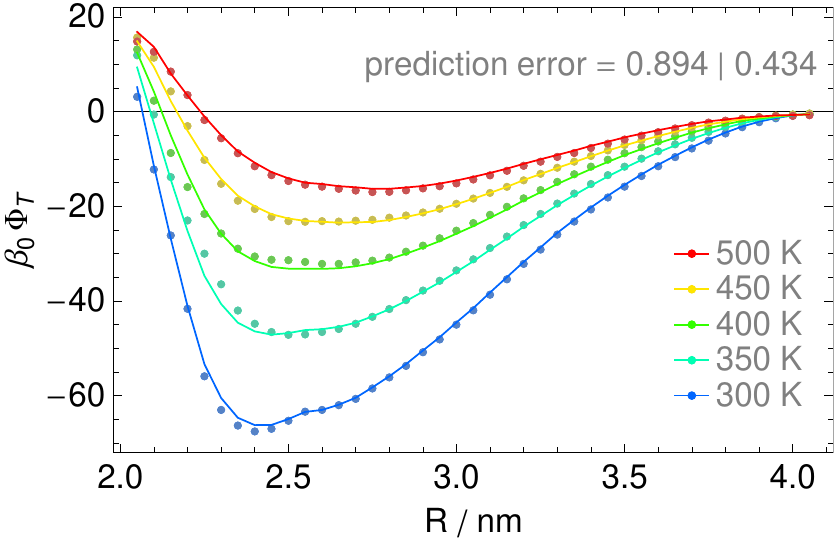}}
\hspace*{1.5em}
\parbox[t]{7cm}{{\Large\sf D}\\[-2ex]
\includegraphics[height=30ex]{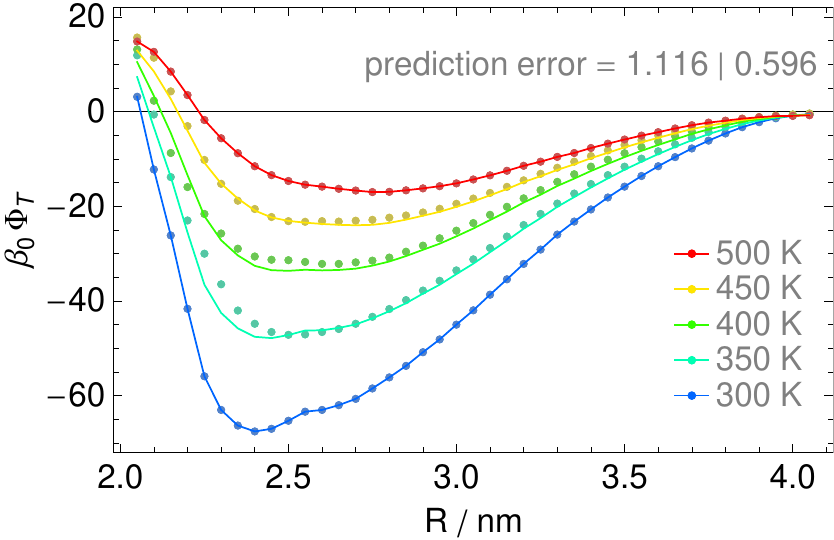}
}
\caption{Prediction of the expansion coefficients.
Panels A and B show the expansion coefficients (dots) obtained from MD simulation data of five temperatures via the pseudo-inverse.
These coefficients fit the PMFs of the MD simulation best (C).
The area (red) around the best fit contains the coefficients obtained for pairs (gray) of temperatures; the border lines represent the envelope of all ten pairs of temperatures and thus indicate the uncertainty for our PMF calculation.
Panel D shows the fit for just two PMF curves (300\,K, 500\,K); this is the result of  \cite{BLGHG14}.
The ``prediction error'' represents the mean square deviation (first value); the second value represents the error restricted to NC-distances $R>R_\text{c}^0=2.35\,\text{nm}$. 
\label{fig2}}
\end{figure}

The coefficients $a_n$ can be calculated using theoretical methods of statistical physics (see Sect.~\ref{sectStatPh} below).
In practice, very few of them need to be known to reproduce the experimental data. For dispersive molecular interactions (e.g., the Lennard-Jones interactions between the molecular groups of our NCs), the first two are sufficient (Fig.~\ref{fig2}AB),
\begin{align}\label{eqHT2}
\beta A=a_0+\beta a_1+\bigO(\beta^2)\,. 
\end{align}

Without experimental or theoretical knowledge, simulations for two \cite{BLGHG14} or more temperatures $\beta_n=1/(k_BT_n)$ are necessary to estimate the two coefficients.
By applying the pseudo-inverse $p=(m^{\rm T}\cdot m)^{-1}\cdot m^{\rm T}$ of the Vandermonde matrix $m=\left((1,\beta_1),(1,\beta_2),\dots\right)^{\rm T}$, 
\begin{align}\label{eqa0a1}
(a_0,a_1)^{\rm T}=p\cdot(\beta_1A_{T_1},\beta_2A_{T_2},\dots)^{\rm T}\,,
\end{align}
we obtain a fit of the coefficients for any given number of simulations
while ensuring that the fitting error $\varepsilon=\beta A-a_0-\beta a_1$ is minimal,
$|\varepsilon^{\rm T}\cdot\varepsilon|\to\min$ (cf.~Fig.~\ref{fig2}C).

For di-polar interactions, second order expansion is
needed, etc., and the number of necessary simulations and coefficients $a_n$ in Eq.~Eq.~\eqref{eqHT2} increases. Again, the best fit is achieved through Eq.~Eq.~\eqref{eqa0a1} in a higher dimension.

\subsection{Interpolation and extrapolation}

Going back to the linear expansion Eq.~\eqref{eqHT2}, let us consider the case of two unknown coefficients. Then the pseudo-inverse reduces to the usual inverse. When inserting the resulting coefficients (cf.~(2.14--15) in \cite{BLGHG14})
into the expansion, 
\begin{align}\label{eqPhiaa}
\beta\Phi_T=\frac{(\beta-\beta_2)\beta_1\Phi_{T_1}+(\beta_1-\beta)\beta_2\Phi_{T_2}}{\beta_1-\beta_2} \,,
\end{align}
one identifies the solution as linear interpolation for $\beta\Phi_T$.
Furthermore, one realizes that in accordance with the high temperature expansion the interpolation parameter is given by the inverse temperature.
The interpolation results are shown in Figure~\ref{fig2}D.

Now let us assume that one coefficient is given. Then only one simulation (performed at temperature $T_0$, say) is needed to calculate $\Phi_T$.
If $a_0$ is known,
\begin{align}\label{eqPhia0}
\Phi_T=(\beta-\beta_0)\,\Delta a_0+\Phi_{T_0}\,,
\end{align}
and if $a_1$ is known,
\begin{align}\label{eqPhia1}
\Phi_T=\left(1-\frac{\beta_0}{\beta}\right)\Delta a_1+\frac{\beta_0}{\beta}\,\Phi_{T_0}\,.
\end{align}
These formulas are readily derived from Eq.~\eqref{eqHT2}; note that the coefficients require the following adjustment, $\Delta a_n(R)=a_n(R)-a_n(\infty)$.

\subsection{Physical particle density}

\begin{figure}
\center
\parbox[t]{7cm}{{\Large\sf A}\\[2ex]
\hspace*{1ex}
\includegraphics[height=26ex]{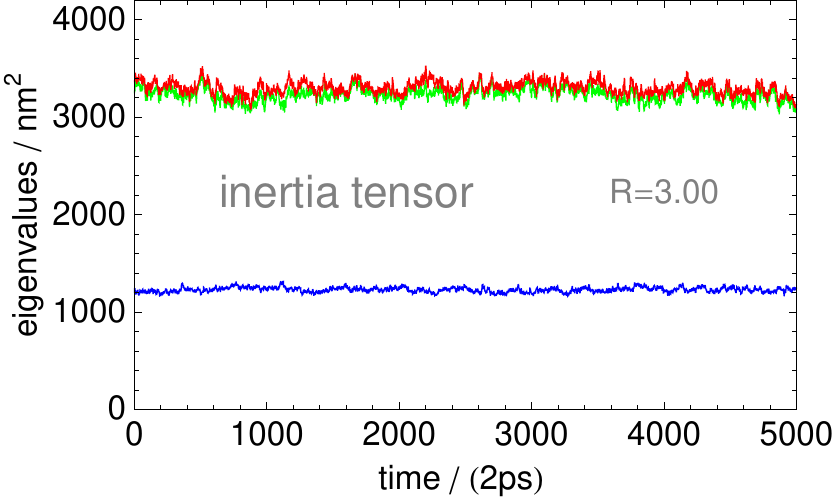}
}\parbox[t]{7cm}{{\Large\sf B}\\[-18ex]
\hspace*{.5ex}
\includegraphics[height=63ex]{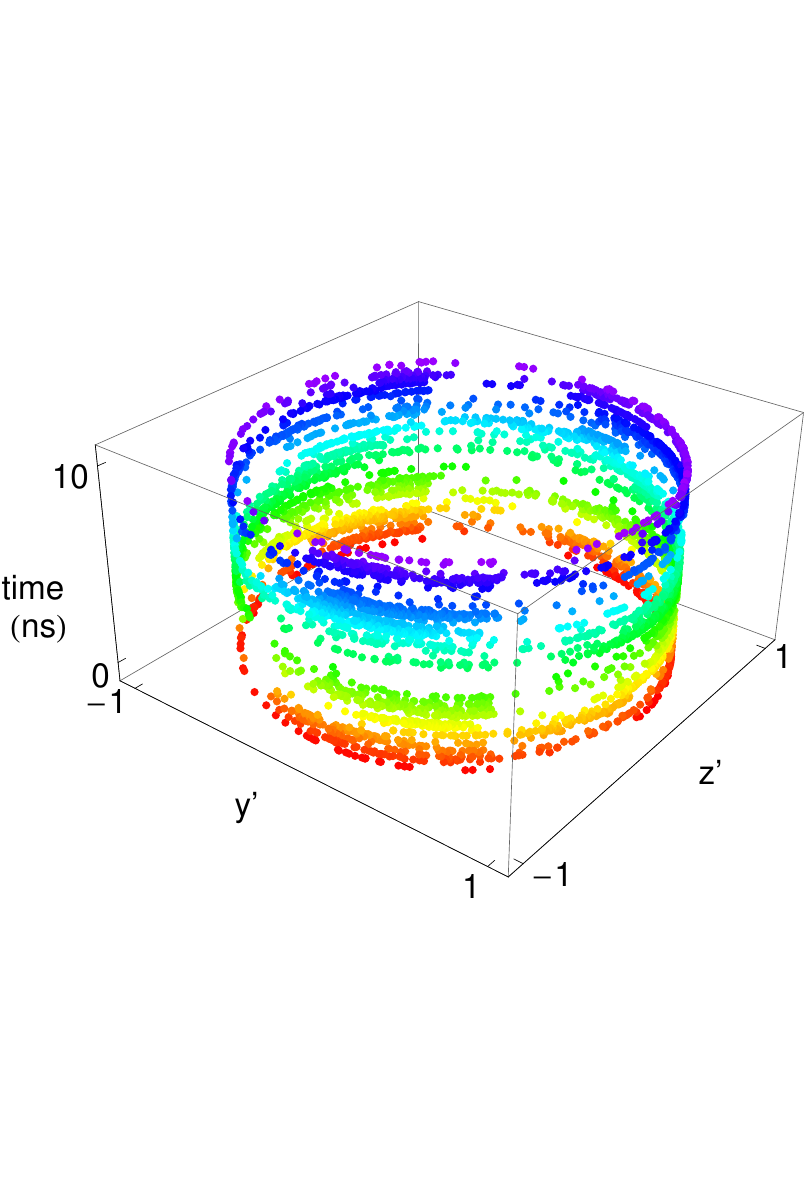}}
\hspace*{1.5em}
\\[-13ex]
\parbox[c]{6.5cm}{{\Large\sf C}\\[-2ex]
\hspace*{7ex}
\includegraphics[height=25ex]{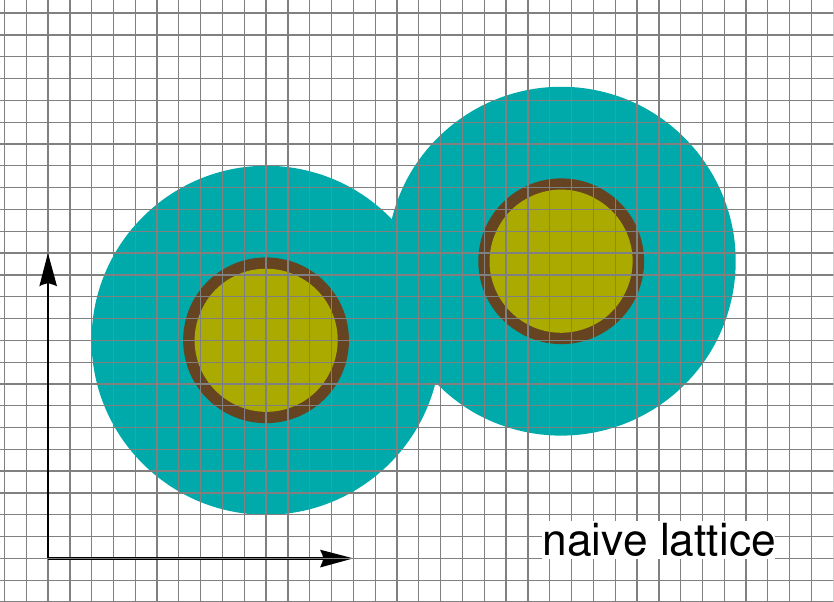}}
\parbox[c]{1.2cm}{\center\sf PCA\\$\Longrightarrow$}
\parbox[c]{6.5cm}{
\includegraphics[height=25ex]{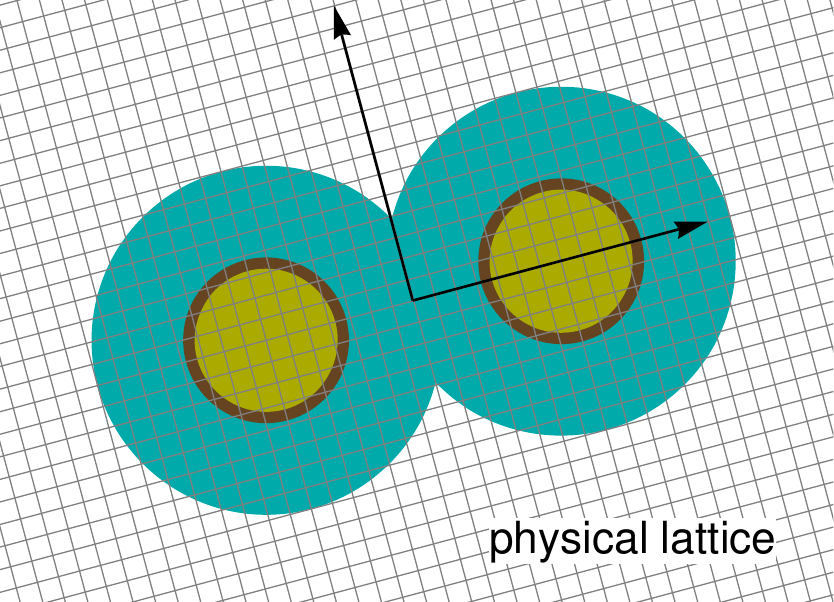}
}\\[4ex]
\caption{PCA-correction. Panel A shows the three eigenvalues of the inertia tensor (for a representative NC-distance of $R=3\,\text{nm}$) after each time step of the MD simulation. The values fluctuate around one small and two (very similar) large numbers. Panel B shows the eigenvectors that correspond the largest eigenvalues over the simulation time; the eigenvectors to the smallest eigenvalue have been aligned to the $x'$-axis (the common $x$-direction); cylindrical symmetry is recognizable.
Panel C illustrates the alignment of the lattice in accordance with the center of mass and the smallest eigenvector of the inertia tensor. This procedure must be applied to the trajectory data after each time step of the MD simulation.  
\label{fig3}}
\end{figure}

To reduce the number of simulations, we will calculate the expansion coefficients in Eq.~\eqref{eqHT2} using three models of density functional theory (Sect.~\ref{sect3}).
By applying this methodology we must determine the density of the relevant interaction sites.
Naive sampling does not account for the momenta that are added to the nano system when coupled to a thermostat.
During each time step $\Delta t$, the stochastic dynamics integrator adds randomly chosen momenta to the interaction sites \cite{VB88}.

The center of mass and the tensor of inertia calculated from our simulation data fluctuate notably (cf.~Fig.~\ref{fig3}A).
Therefore we propose a procedure that corrects for the unphysical interference
(cf.~Fig.~\ref{fig3}C). We re-adjust the center of mass and the tensor of inertia of our trajectory data by applying principal component analysis (PCA) after each time step of a simulation run.
That is, we move the lattice (Fig.~\ref{fig3}C)
so that effectively we sample a density (referred to as physical)
that keeps the center of mass and the tensor of inertia (i.e., because of rotational symmetry only the lowest eigenvalue) of the nano system fixed.
Here averages over the 5000 time steps of a simulation run serve as reference values.
In our simulations the gold cores are kept frozen in space, therefore we only correct the data of the ligand caps.

Performing PCA (using the LAPACK software package \cite{ABBBD90}), we observe one small- and two very similar large eigenvalues (Fig.~\ref{fig3}A).
The eigenvectors of the two large eigenvalues form an almost steadily distributed circle around the $x$-axis (cf.~Fig.~\ref{fig3}B).
Therefore it is reasonable to propose cylindrical symmetry for the density of the interaction sites in the cap, even though the cores of the NCs are icosahedras \cite{WTD05}. This provides another, practical argument why performing our simulation by freezing the NCs' cores in space (and not only restraining their centers of mass and allowing for rotations) is an acceptable simplification \cite{BLGHG14}.

\subsection{Perturbation theory \label{sectStatPh}}

The coefficients of the high temperature expansion Eq.~\eqref{eqHT1} are determined by the molecular interaction.
We recall that the partition function
$Q=(\Lambda^{3N}N!)^{-1}\int\exp(-\beta V^{(N)})dr^N$ defines
the Helmholtz free energy
\begin{align}
A=-\beta^{-1}\ln Q
\end{align}
in terms of the molecular potentials $V^{(N)}=\sum_{i<j\le N} v^{(ij)}(|r_i-r_j|)$; $N$ denotes the number of interaction sites, $v^{(ij)}(\cdot)$ the pair potential between sites $i$ and $j$, and $\Lambda$ the thermodynamic wavelength.
Here, perturbation theory, which model the transition of the free energy from a well-known reference fluid (with pair potential $v_0$) to a target fluid (with pair potential $v$), defines the coefficients $a_n$, $n>0$ \cite{HM06}.
The high-temperature expansion, Eq.~\eqref{eqHT1}, reads
\begin{align}
\beta A=a_0+\beta\,\langle W\,\rangle+\bigO(\beta^2)\,,
\end{align}
where $W=V^{(N)}-V^{(N)}_0$ represents the perturbation and $\langle W\rangle=\int W\exp(-\beta V^{(0)}_0)\,dr^N/\int\exp(-\beta V^{(0)}_0)\,dr^N$
denotes the corresponding average for the canonical ensemble of the reference fluid.
Examples are content of the subsequent section.

\section{Results} \label{sect3}

Three thermodynamic models will 
be used to calculate one of the expansion coefficients in Eq.~\eqref{eqHT1}. Relying on linear expansion Eq.~\eqref{eqHT2}, MD simulations for only one reference temperature (e.g., $T_0=\text{350\,K}$) are needed to determine the PMF for any other temperature $T$; see Eq.~\eqref{eqPhia0} and Eq.~\eqref{eqPhia1}.
We reconstruct PMFs for the available set of simulations
and explain the intricacies when applying the models.
We start off with the simplest one.

\subsection{The ideal gas model}

Here we assume that the entropic contribution
(i.e., the coefficient $a_0=-S/k_B$)
represents an ideal gas.
It is, however, the ideal gas contribution of a heterogeneous fluid with a density distribution obtained from the MD simulation.
Given a continuous particle density, $\rho=\lim\frac{\Delta N}{\Delta V}$, as $\Delta V\to0$,
the ideal gas contribution is known (for classical fluids) as
\begin{align}
a_0
=\int\rho(r)\left(\ln\left(\Lambda^3\rho(r)\right)-1\right)\,d^3r\,.
\end{align}
This formula can be simplified when only taking differences of the Helmholtz free energy, \begin{align}\label{a0IG2}
a_0
=\int\rho(r)\ln\rho(r)\,d^3r\,.
\end{align}
According to Eq.~\eqref{eq:PhiR},
the difference concerns two states: one, where the NCs are located at a distance $R$ apart from each other and, another, where one of two NCs is separated to infinity.
We thereby assume that the intra-molecular partition sum is unchanged for the two states modeled with Eq.~\eqref{a0IG2}.

The result represents Shannon's entropy \cite{S48}. Its only input is the particle density, which can be obtained from trajectory data.
While considering differences of the free energy, we restrict ourselves to interaction sites that markedly change the position with respect to (the center of mass of) their NC when moving the NCs in space. These interaction sites are the $\text{CH}_x$-groups (i.e., $\text{CH}_2$- and $\text{CH}_3$-groups) of the thiol chains.

\begin{figure}
\center
\parbox[t]{7cm}{{\Large\sf A}\\[-2ex]
\hspace*{.5em}
\includegraphics[height=30ex]{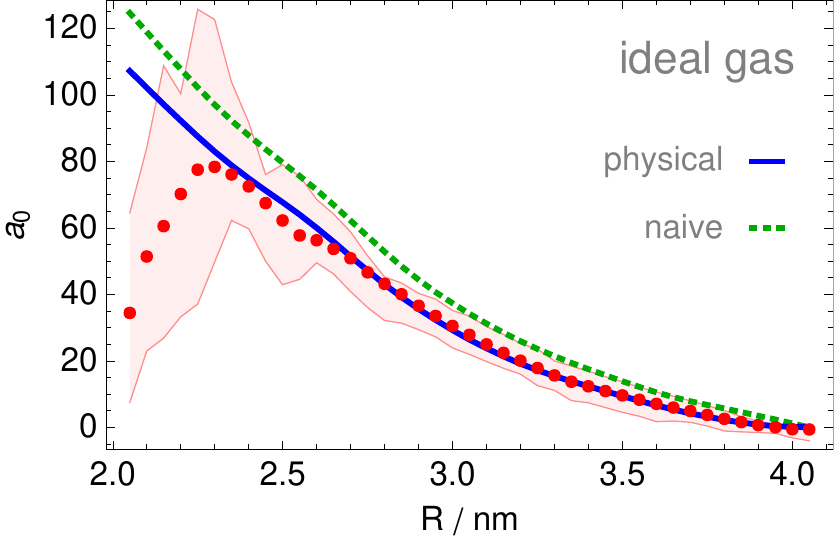}}
\hspace*{1.5em}
\parbox[t]{7cm}{{\Large\sf B}\\[-2ex]
\includegraphics[height=30ex]{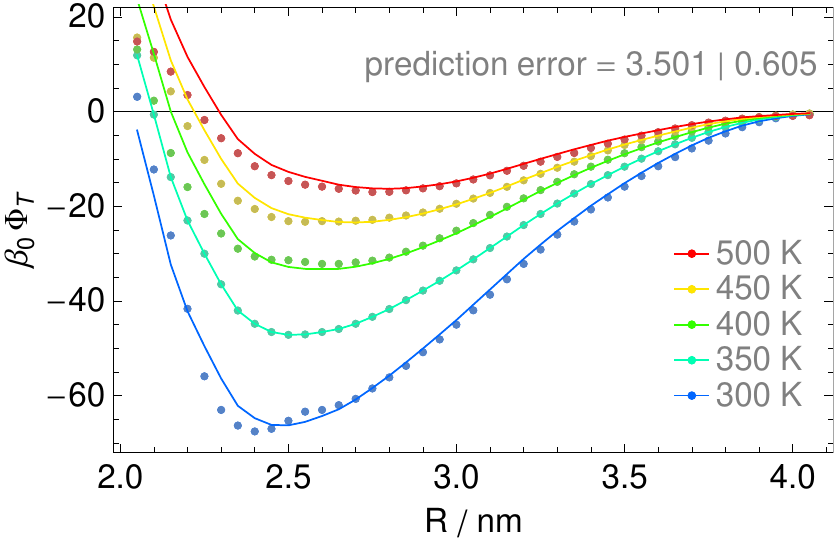}}\\[2ex]
\parbox[t]{7cm}{{\Large\sf C}\\[-2ex]
\hspace*{.5em}
\includegraphics[height=30ex]{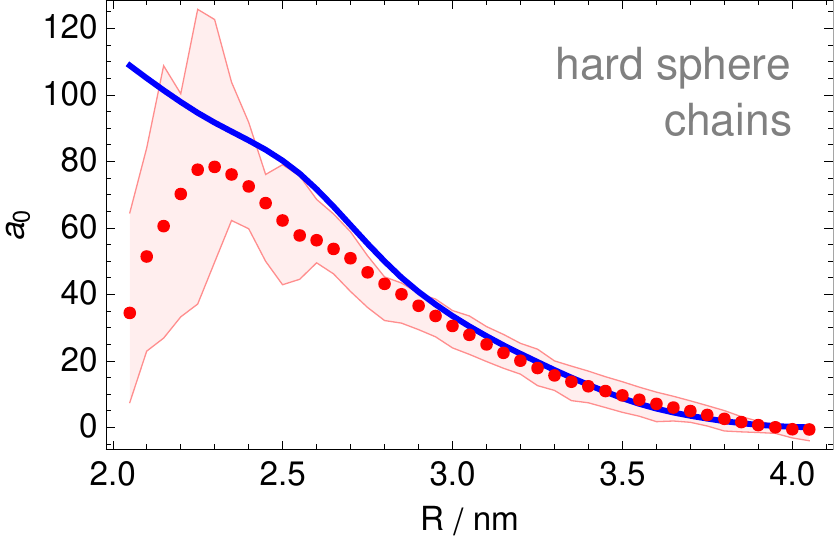}}
\hspace*{1.5em}
\parbox[t]{7cm}{{\Large\sf D}\\[-2ex]
\includegraphics[height=30ex]{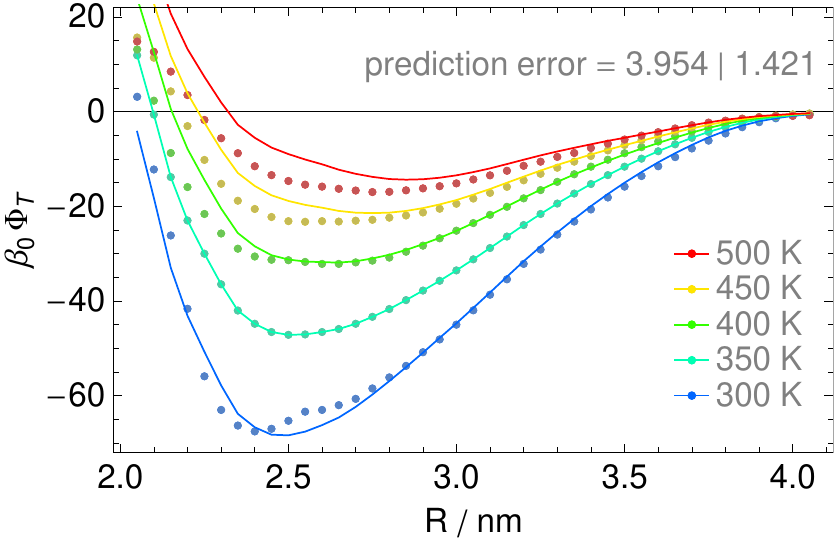}}\\[2ex]
\parbox[t]{7cm}{{\Large\sf E}\\[-2ex]
\includegraphics[height=30ex]{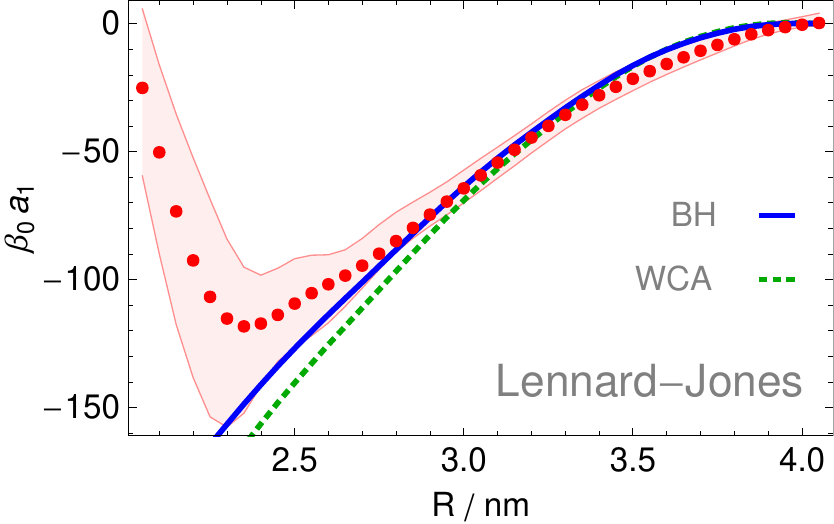}}
\hspace*{1.5em}
\parbox[t]{7cm}{{\Large\sf F}\\[-2ex]
\includegraphics[height=30ex]{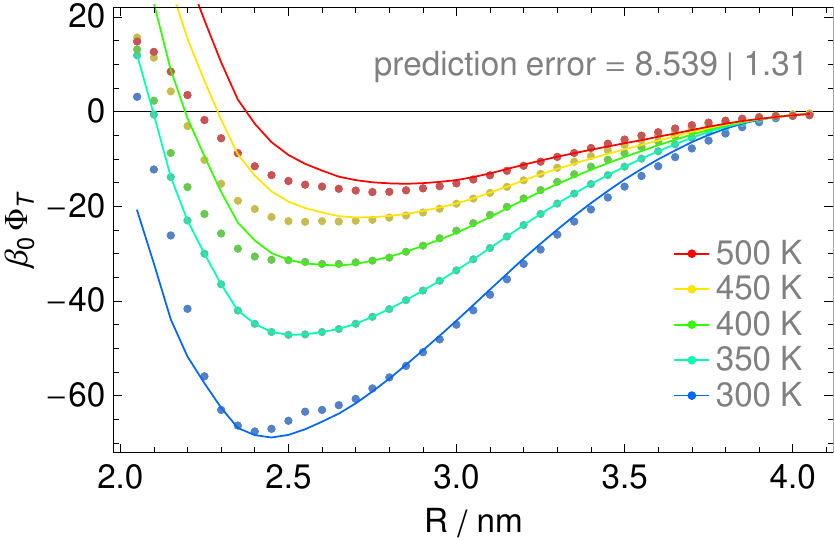}}
\caption{Comparison of PMFs from MD simulations and PMFs predicted from three models:
the ideal gas (AB), the hard-sphere chains (CD), and the Lennard-Jones fluid (EF). Panels AC and E show the coefficients $a_0(R)$ and $a_1(R)$, resp.
Panels BDF display the PMF curves
predicted from the models, that are represented by blue lines in the
corresponding diagrams ACE.
For the ideal gas method (A), the coefficients relying on a naive particle density (green, dotted) are up to 20\% larger than the coefficients based on the physical particle density (blue).
Above a critical distance ($R_\text{c}^0=2.35\,\text{nm}$), the blue coefficient curves approximate the expected coefficients (red dots) very well.
Below the critical distance the deviation becomes large.
For the two other methods, the critical distance is higher ($R_\text{c}^1=2.7\,\text{nm}$).
For the Lennard-Jones fluid, BH's approach (blue) performs better than WCA's (green, dotted).
For all methods, the reference temperature $T_0=350\,\text{K}$ is used.
Fluctuations in the calculated coefficient curves ($<5\%$) were smoothed out.
\label{fig4}}
\end{figure}

The simplicity of this approach is deceiving.
The density in our formulas is a continuous quantity, but the 
positional coordinates of the interaction sites obtained from an MD simulation are determined for discrete time steps.
We find that a cubic lattice with mesh size $\delta=\text{.05\,nm}$ allows for sampling a reliable discrete density, leading to reproducible values for the Helmholtz free energy. Yet these values turn out to be ca.~20\% too large compared to the predicted values 
(cf.~Fig.~\ref{fig4}A).

We believe the discrepancy is due to the thermostat. 
The stochastic dynamics integrator used in the simulations randomly adds momenta to the molecular groups. This is exactly what we observe from the trajectory data.
The center of mass and the tensor of inertia of the interaction sites fluctuate over time (Fig.~\ref{fig3}AB), despite the fact that the molecular groups of the two NCs are the only particles simulated in space.

To correct for the unaccounted physical interaction we utilize a density obtained through PCA,
ending up with an entropic contribution that is significantly smaller than the one obtained from naive sampling.
Above a certain distance of the NCs, $R>R_\text{c}^0=2.35\,\text{nm}$, the
agreement of our prediction with the $a_0$-data (as determined from MD simulations at various temperatures) is very good (Fig.~\ref{fig4}A).

In order to appreciate this result, we note that an accurate prediction of $a_0$ (as shown in Fig.~\ref{fig4}A) leads to a reliable estimate of the
PMF for any temperature (via Eq.~\eqref{eqPhia0}) using MD simulation data of only one temperature. The predicted PMF at various temperatures are shown in Fig.~\ref{fig4}B.

\subsection{Reference fluid of hard-sphere chains}

A much more advanced model for the thiol ligands of our NCs is given by Wertheim's theory of hard-sphere chains \cite{W87}.
Here the entropic contribution represents a chain with $m$ hard-sphere segments, containing Carnahan-Starling's term \cite{CS69}.
In contrast to the previous approach (ideal gas model),
the ideal gas term is neglected \cite{GHL04} here,
because now the
reference ideal gas state is one where interaction sites
are considered free (unconstrained to chain structure).
The ideal gas contribution then cancels in the difference
of Eq.~\eqref{eq:PhiR}.
The reference fluid (term of zero order) is
\begin{align}
a_0
=\int\overline\rho\left(\frac{4\eta-3\eta^2}{(1-\eta)^2}-\left(1-1/m\right)\ln\frac{1-\eta/2}{(1-\eta)^3}\right)d^3r\,,
\end{align}
where $\eta=V_\sigma\overline\rho$ denotes the packing fraction and $\sigma$ the hard-sphere diameter.
According to Tarazona \cite{T84}, a weighted particle density
\begin{align}\label{eq:rhobar}
\overline\rho(r)=\int\rho(r')\theta(\sigma-|r-r'|)\,dr'\,
\end{align}
is utilized, computed from the physical density $\rho(r)$.

To determine the two parameters, $m$ and $\sigma$, we use a homo-segmented group contribution method \cite{SSG14}, where the thiol chains are composed by a fractional number ($m=3.804$, by~Eq.~2 in \cite{SSG14}) of associating hard-sphere segments with equal diameter ($\sigma=.386\,\text{nm}$, by~Eq.~3 in \cite{SSG14}) and hence equal volume ($V_\sigma=\pi\sigma^3/6$).
Figure \ref{fig4}CD shows the results. Similar to the ideal gas model, the fit is very good for distances of the two NCs above
$R_\text{c}^1=2.7\,\text{nm}$.

Lo Verso et al.~\cite{LYEB11} also used a chain model for a similar but artificial nano system, where the density $\rho(r)$ was sampled naively. In contrast to our ideal gas model, no involvement of the thermostat and no dependence on the mesh size has been reported.
In fact, this coincides with our observations. When applying the naive density to our chain model we obtain almost the same results as when applying the physical density.
This could be due to the weighted particle density and its smoothing behavior. 
One has to be careful though when implementing Eq.~\eqref{eq:rhobar} on a lattice; the lattice must be fine enough to sufficiently represent a single hard-sphere volume $V_\sigma$ by summation over its cells.

\subsection{Lennard-Jones fluid}

In our third model we
take a different approach. Rather than estimating the reference fluid $a_0$, we now estimate the $a_1$ term (via Eq.~\eqref{eqPhia1}). The approach is similar in practical application, because the respective other term ($a_0$ or $a_1$) is determined by MD simulations and Eq.~\eqref{eqHT2}. 
We consider the ligand segments as a Lennard-Jones fluid. That is, we apply the force field that was used in the MD simulation, $v(r)=4\epsilon\left((\sigma/r)^{12}-(\sigma/r)^{6}\right)$.
However, we only consider the $\text{CH}_x$-groups as interaction sites,
and we only compute the interaction term in first order perturbation theory,
\begin{align}\label{eqa1LJ}
a_1
&=\frac{1}{2}\iint\rho(r)\rho(r') g(|r-r'|)v_1(|r-r'|)d^3rd^3r'\,,
\end{align}
where $v_1=v-v_0$ denotes the perturbation and $g(\cdot)$ the pair distribution of the reference fluid.
As reference fluid we consider hard-spheres, and numerically we approximate their pair distribution using the Percus-Yevik's method \cite{PY58}.
We apply two standard theories: the one by Barker-Henderson (BH) \cite{BH67} and the one by Week-Chandler-Andersen (WCA) \cite{WCA71}.
The density $\rho(r)$ is sampled again from the simulated trajectories after PCA is performed.
The results are shown in Figure \ref{fig4}EF. 
The fitting is very good above the distance $R_\text{c}^1=2.7\,\text{nm}$.
Naive sampling produces slightly worse results.

BH performs better than WCA. This is a bit surprising, as BH's higher order terms --- unlike WCA --- usually also contribute to Eq.~\eqref{eqHT1}. Maybe the choice of the reference fluid with an only temperature dependent hard-sphere diameter is responsible. BH's reference potential is steeper than WCA's and thus closer to a hard-sphere fluid.

The computational effort of summing over all interactions Eq.~\eqref{eqa1LJ} is 
much higher than in the previous two methods.
We took advantage of cylindrical and mirror symmetry when sampling the particle density, reducing the computation time by the factor $1/64$.

\section{Discussion}

For an exemplar system of two thiol capped gold core NCs we estimated the PMF using three models.
Above a critical distance ($R>R_\text{c}^x$), each of the three models led to PMF curves of satisfactory quality.
The methods underlying these models differ considerably with respect to the number of input parameters and the computational effort.

\subsection{Model comparison}

The ideal gas model returned the best PMFs and requires no extra parameters. Yet one must correct the simulated particle distribution via PCA to obtain a physical density. As MD simulations usually involve (virtual) interaction with a thermostat,
our correction should be relevant for many similar problems.
One disadvantage of this methodology is the additional computational cost (when performing PCA for 5000 data sets), but it is far less than the cost of possibly additional MD simulations.

The hard-sphere chain model requires knowledge of two additional parameters, the hard-sphere diameter and the chain length. Both these parameters were obtained independently from a homo-segmented group contribution method.
The fit of the PMF-data is almost as good as with the ideal gas model.
For further improvement, one may try to employ a hetero-segmented method \cite{SSG14}.
Because of the weighted density that enters the formulas, the model turned out to be robust with respect to small imprecisions of the particle density.
That is, compared to the ideal gas model, no PCA correction is needed.
Therefore, the computational cost is smaller too.

The 'Lennard-Jones model', did not perform as well as the other two models, in particular for small NC-distances.
There are several possible reasons.
The radial pair distribution function for spheres is only a rough approximation for the pair distribution of heavily structured NCs \cite{TB05}.
For densely overlapping ligands (i.e., NC-distances $R\lesssim2.7\,\text{nm}$), the sum over the Lennard-Jones interactions turned out too high.
Furthermore, we introduced a lattice of rather big mesh size, because summation over all the attractive forces is costly and extends with cubic power.
The proposed rotational and mirror symmetries for the lattice entries will have contributed to errors as well.

In each of the three models, the NC ligand layer is studied as a fluid. 
Taking this aspect of the approach seriously and
regarding the ligand layer as a solvent for the core, we are able to clarify the role of an additional force observed at small distances of the NCs, as explained now.

\subsection{Model extension: depletion attraction}

\begin{figure}
\center
\parbox[t]{7cm}{{\Large\sf A}\\[-2ex]
\includegraphics[height=30ex]{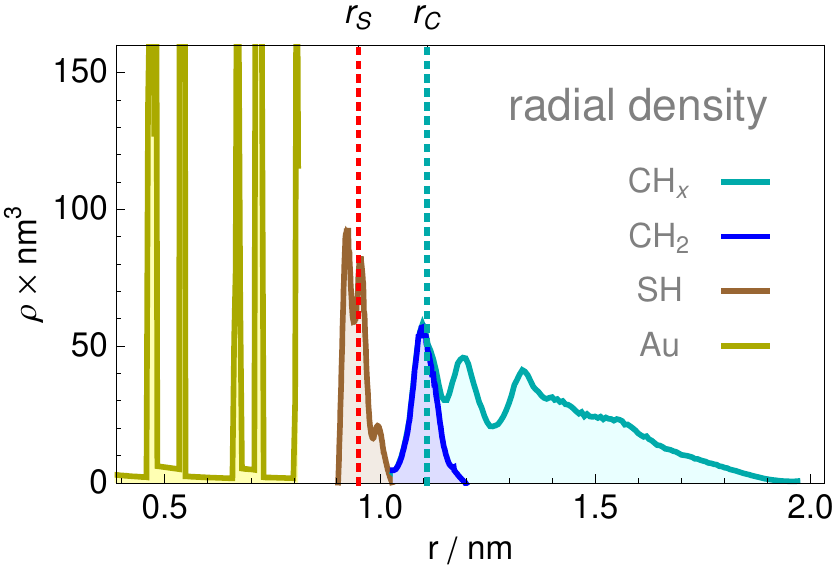}}
\hspace*{1.5em}
\parbox[t]{7cm}{{\Large\sf B}\\[-2ex]
\includegraphics[height=30ex]{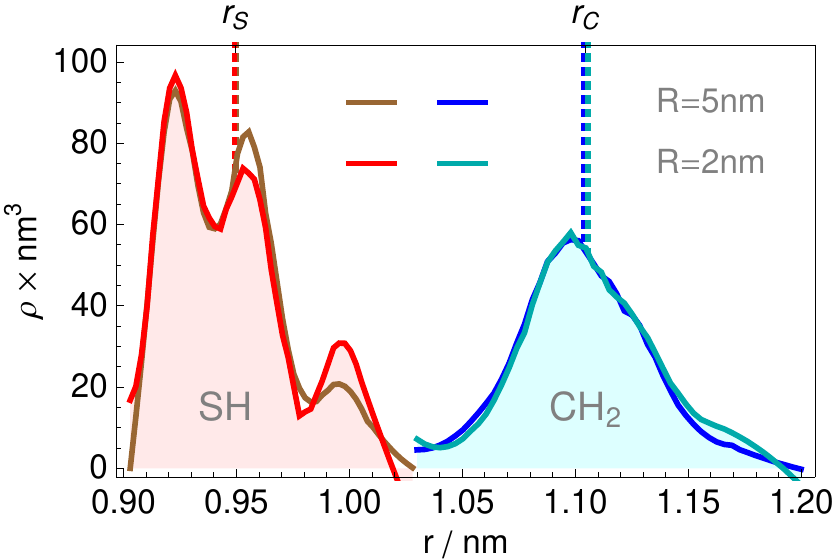}
}
\caption{Radial density. Density profiles of NC\,1 und NC\,2 practically coincide. In Panel A, the NCs are separated by a distance of $R=5\,\text{mn}$.
For smaller separation distances (cf.~B) the densities have shifted --- away from intermediate radial distances, for the $\text{SH}$-group,
and towards larger distances, for the first $\text{CH}_2$-group; that is, $\text{SH}$ makes room for secondary $\text{CH}_2$-groups of the other NC.
\label{fig5}}
\end{figure}

\begin{figure}
\center
\parbox[t]{7cm}{{\Large\sf A}\\[-2ex]
\hspace*{.5em}
\includegraphics[height=30ex]{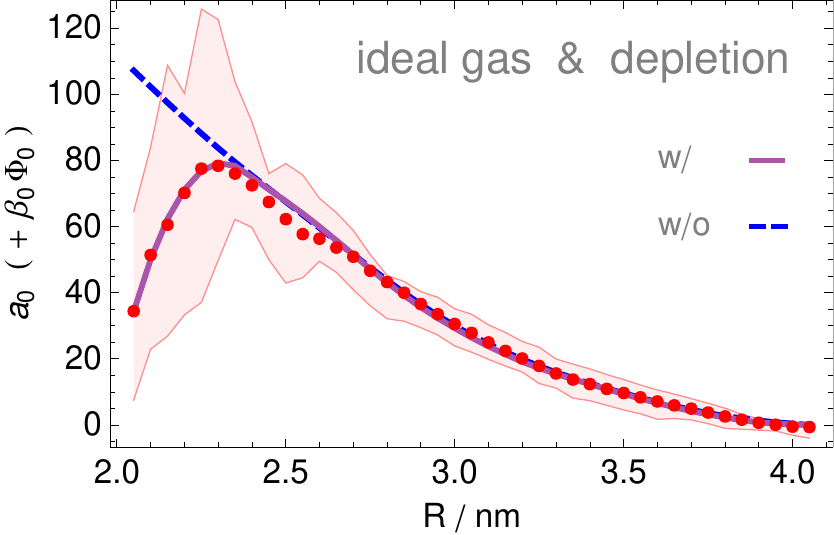}}
\hspace*{1.5em}
\parbox[t]{7cm}{{\Large\sf B}\\[-2ex]
\includegraphics[height=30ex]{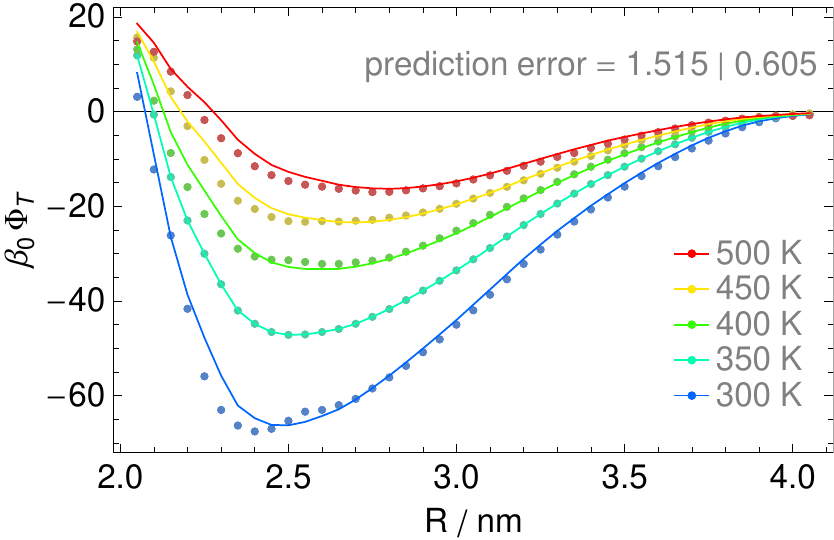}
}\\[2ex]
\parbox[t]{7cm}{{\Large\sf C}\\[-2ex]
\includegraphics[height=30ex]{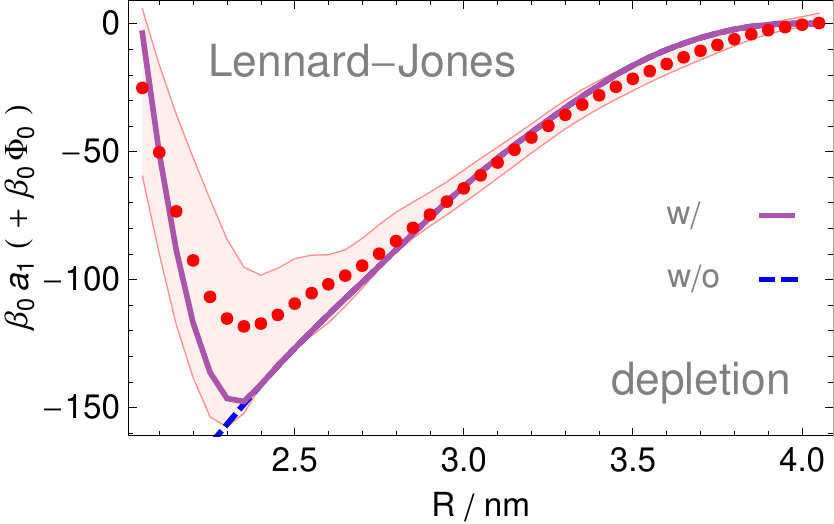}}
\hspace*{1.5em}
\parbox[t]{7cm}{{\Large\sf D}\\[-2ex]
\includegraphics[height=30ex]{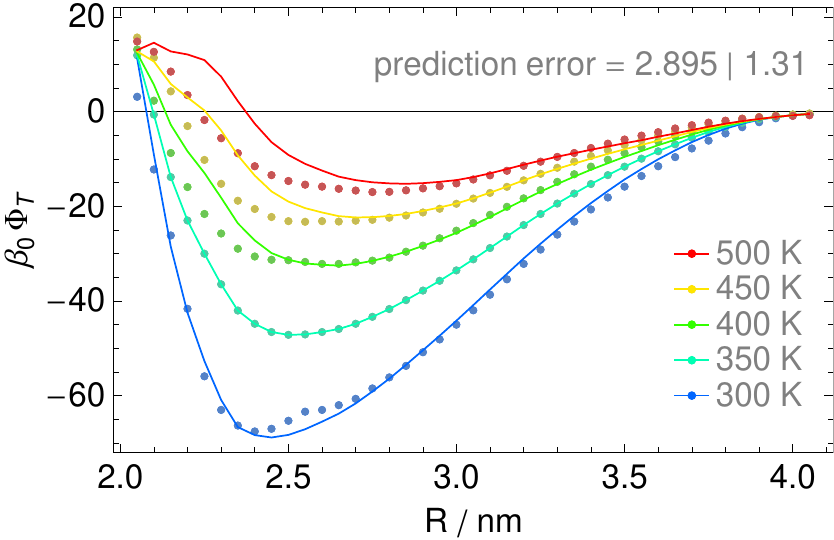}
}
\caption{Effect of depletion forces on the PMF. Expansion coefficients and PMFs of the ideal gas model and the Lennard-Jones model are shown with and without contribution due to depletion. The onset distance is estimated by theoretical arguments and thus chosen to be the same for both model fluids, $R_0=2.357\,\text{nm}$. This value is very close to the best fit obtained for the ideal gas. The parameter $\kappa$ is fitted to each model, individually.
\label{fig6}}
\end{figure}

The change in monotony of the entropic coefficient $a_0(R)$
around the critical distance $R=R_\text{c}^0$
and the relatively small error bounds (cf.~Fig.~\ref{fig1}A)
suggest that there is an additional attractive force $F_0(R)$ not yet accounted for in our models.
Let us
assume that this force behaves according to Hook's law with spring constant $\kappa$ --- with onset at distance $R_0\gtrsim R_\text{c}^0$ and becoming larger at smaller distances of the two NCs. The additional contribution to the PMF would then be quadratic,
\begin{align}\label{eqPhi0}
\beta\Phi_0(R)=-\frac{\kappa}{2}(R_0-R)^2\,\theta(R_0-R)\,,
\end{align}
with $\theta$ being the Heaviside function. 
We adjusted the parameters, $R_0$ and $\kappa$, to the PMF-data generated by the MD simulation.
The best result is achieved for a onset distance $R_0=2.374\,\text{nm}$ slightly above the critical distance $R_\text{c}^0=2.35\,\text{nm}$;
the regression shows excellent agreement with the data
(Fig.~\ref{fig6}A), with error values similar to \cite{BLGHG14}.
The harmonic ansatz function, Eq.~\eqref{eqPhi0}, represents the simplest form of a depletion force \cite{AO54,GED98}.
In practice, adjusting the parameter $kappa$ requires one additional MD simulation. The additional simulation is in conflict with our objective to reduce the simulation effort to a single MD run, and for many applications it is sufficient to simply neglect the short-ranged potential, Eq.~\eqref{eqPhi0}.
Nonetheless, from a more fundamental viewpoint we are interested in elaborating the force further, showing that the short-ranged force contribution is indeed likely the result of a depletion force. In doing so, we show that the distance parameter $R_0$ can be estimated from geometric data.

Depletion forces between two NCs are apparently attractive forces. The effect is caused by the depletion of smaller entities (here interaction sites of the ligands) between the NCs due to geometrical size exclusion. As a result, the local density of ligands between the NCs is taking on small values, leading to a low local pressure. The depletion forces are thus attractive.
Depletion has been used to explain phenomena in the context of nano-sized particles --- with polymer parts taking over the role of solvent particles \cite{KCSZ99,PW02,EB12,MSBNe13}.

In our nano system, only the $\text{CH}_x$-groups move
and thus mimic solvent particles.
Therefore, depletion attraction is supposed start at (or slightly above)
the distance where only one $\text{CH}_2$-group fits in between the two NCs.

The considered $\text{CH}_2$-group is bound to a $\text{SH}$-group, which
is bonded rather immobile
to the gold core of one NC (\#1, say). The surface of the other gold core (of NC\,2) is densely covered with $\text{SH}$-groups (cf.~Fig.~\ref{figA1}B).
The average distance of the two NCs, $R_\text{SC)(S}$, where the first $\text{CH}_2$-group of NC\,1 touches the $\text{SH}$-group surface of NC\,2, should then be an estimate for the onset distance $R_0$.
In fact, we found that numerically the optimal onset distance coincides with
\begin{align}
 R_\text{SC)(S}&=r_\text{C}+\delta_\text{C)(S}+r_\text{S}
 =2.357\,\text{nm}
\end{align}
up to $17\,\text{pm}$ (which is far below mesh size $\delta=50\,\text{pm}$);
here $r_S=.950\,\text{nm}$ and $r_C=1.107\,\text{nm}$ are the mean radii of the $\text{SH}$- and the first $\text{CH}_2$-groups about the gold cores (obtained from trajectory data; cf.~Fig.~\ref{fig5}),
and $\delta_\text{C)(S}=.312\,\text{nm}$ represents the average distance between the centers of $\text{CH}_2$ (NC\,1) and $\text{SH}$ (NC\,2) over the  bumpy $\text{SH}$-surface of NC\,2  (calculated in Appendix~\ref{secA}).

\subsection{Outlook}

\begin{figure}
\center
\parbox[t]{7cm}{{\Large\sf A}\\[-2ex]
\includegraphics[height=30ex]{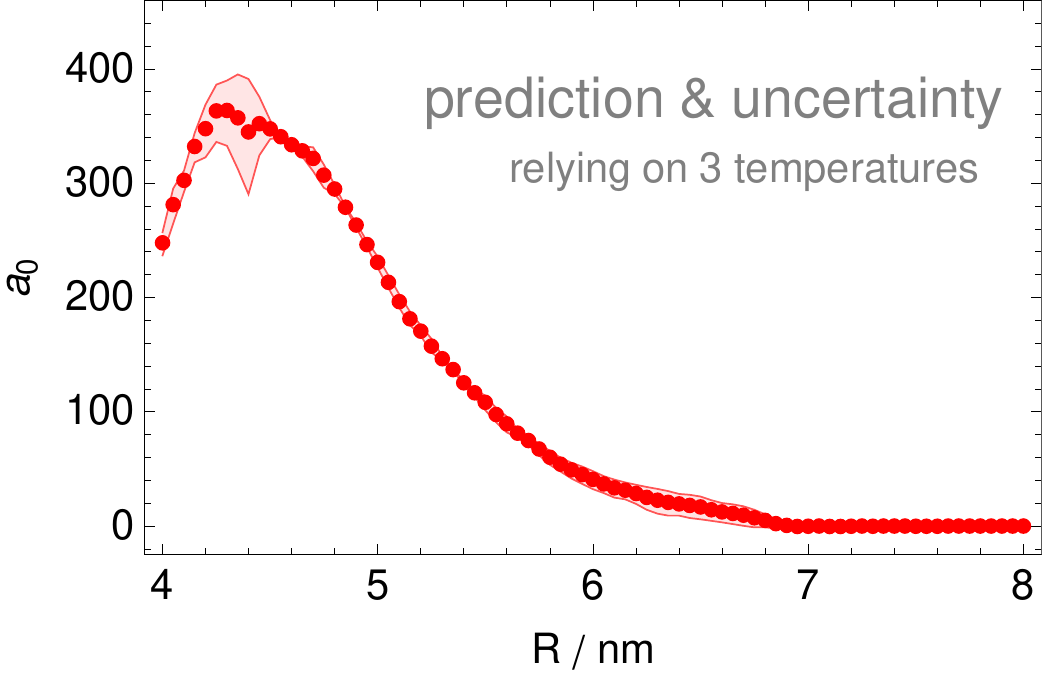}}
\hspace*{1.5em}
\parbox[t]{7cm}{{\Large\sf B}\\[-2ex]
\includegraphics[height=30ex]{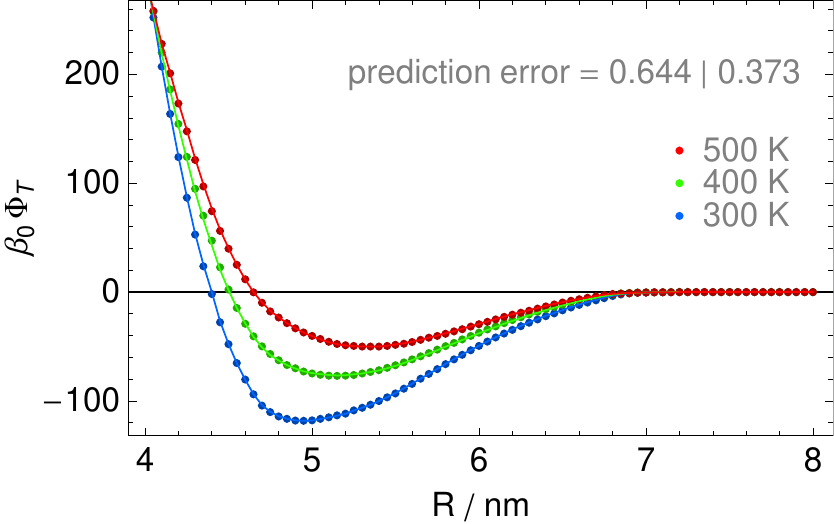}
}\\[2ex]
\parbox[t]{7cm}{{\Large\sf C}\\[-2ex]
\includegraphics[height=30ex]{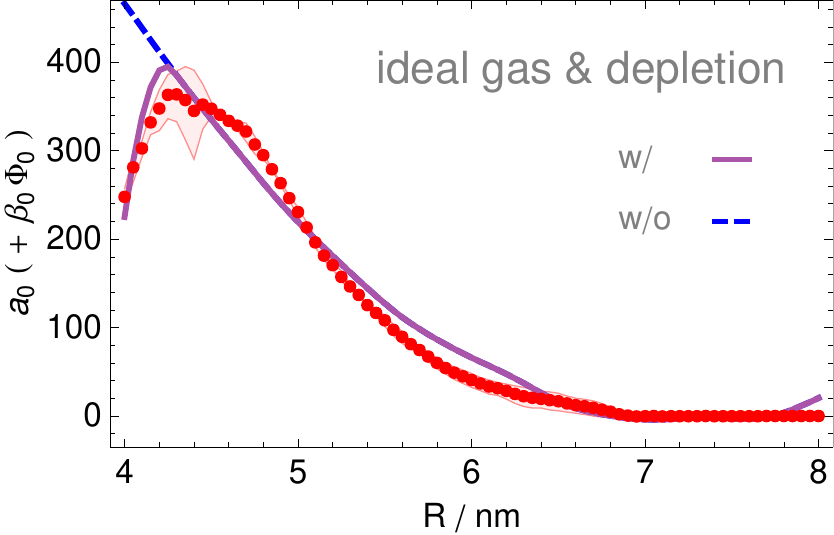}}
\hspace*{1.5em}
\parbox[t]{7cm}{{\Large\sf D}\\[-2ex]
\includegraphics[height=30ex]{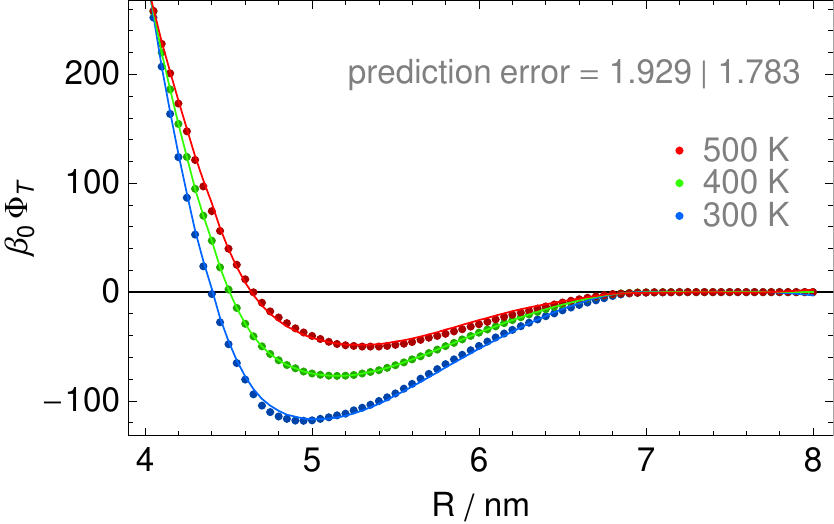}
}
\caption{Modeling results for another pair of NCs, $\text{Au}_{1415}(\text{SC}_{12})_{242}$.
The NCs consist of gold cores with 1415 Au atoms and 242 alkanethiol ligands (one $\text{SH}$- and twelve $\text{CH}_x$-groups). Panel A shows the coefficient $a_0(R)$ based on MD data for three temperatures; the best estimate is indicated by red dots, the corresponding error margin by the surrounding red area. Panel B compares the predicted PMF (lines) and the PMF
obtained from MD simulations (dots). Panels C and D show the results of the ideal gas model with (and without) depletion; $T_0=400\,\text{K}$. The onset distance ($R_0=4.284\,\text{nm}$) has been chosen according to our geometric analysis, based on the mean radii ($r_\text{S}=1.894\,\text{nm}$, $r_\text{C}=2.052\,\text{nm}$) obtained from the trajectory data. Here, the onset distance causing the best fit (4.445\,nm) is notably higher and likely to be attributed to the longer ligands. 
\label{fig7}}
\end{figure}

To test the range of validity we performed MD-simulations for pairs of NCs with other core sizes and ligand lengths, and we also studied heterogeneous NC pairs \cite{BLGHG14}. For all these cases the modeling approach is in good agreement with a series of independent MD results (cf.~Fig.~\ref{fig7}).
It needs to be examined, however, if for more extreme dimensions (e.g., very long ligands) and for other core and/or cap materials our methodology leads to comparably good results.

After two-body interaction, three-body interaction must be studied \cite{BGLHG14}.

\appendix
\section{Depletion: the onset distance \label{secA}}

\begin{figure}
\center\vspace*{-0ex}
\parbox[t]{7cm}{{\Large\sf A}\\[-16ex]\hspace*{-2em}
\includegraphics[height=65ex]{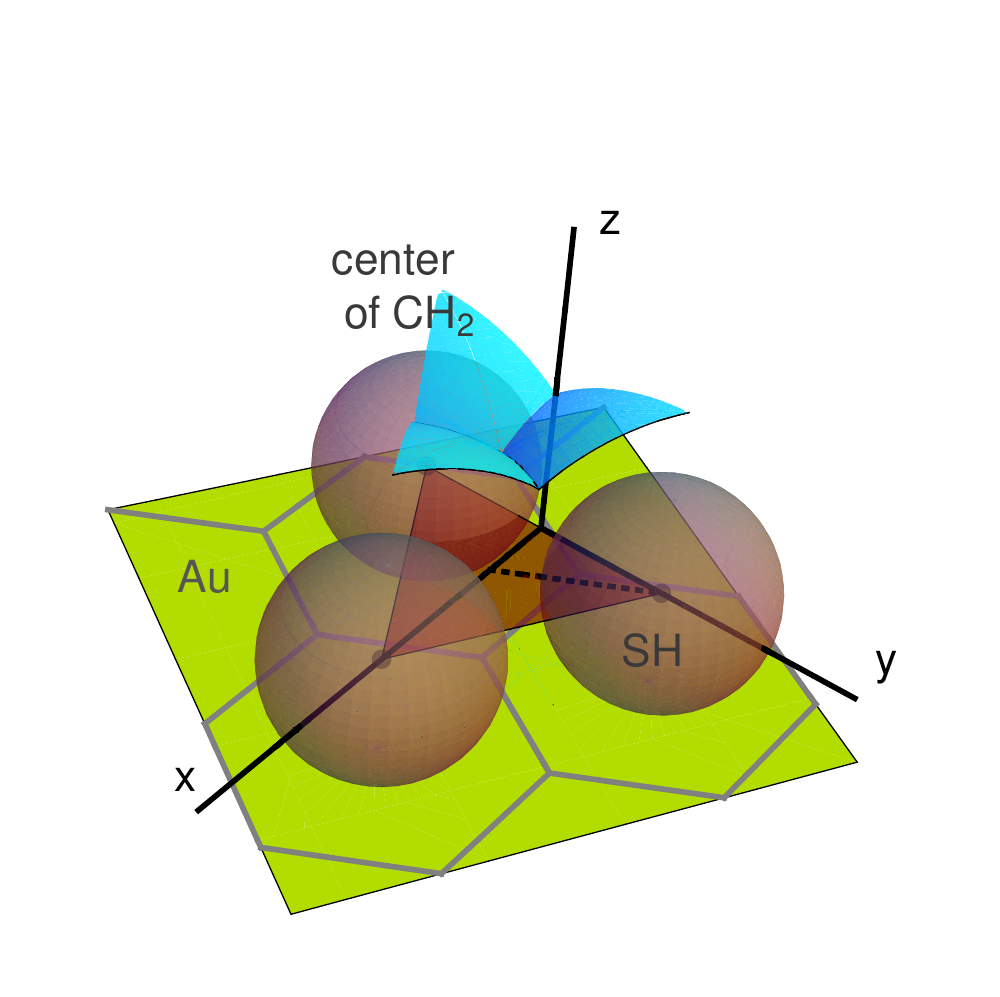}
}
\hspace{1.3cm}
\parbox[t]{7cm}{{\Large\sf B}\\[-2ex]
\hspace*{1em}
\includegraphics[height=44ex]{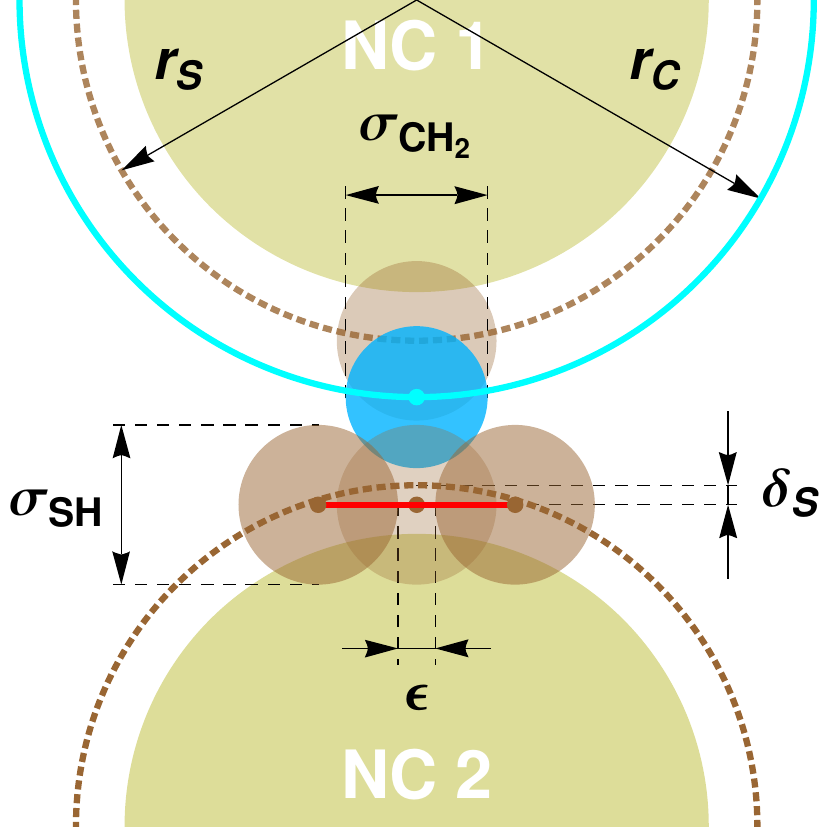}}\vspace*{-4ex}
\caption{Contact distance between two NCs. The blue surface $z(x,y)$ in Panel A illustrates the center of the first $\text{CH}_2$-group of NC\,1 when touching (at least) one of three $\text{SH}$-groups of NC\,2 with centers at zero-level (red triangle). Panel B shows the geometric parameters involved.
\label{figA1}}
\end{figure}

Looking at the geometric structure of the NCs, one cannot recognize a simple rule that determines the positions of the $\text{SH}$-groups adsorbed to the $\text{Au}$-icosahedra.
It is known, however, that
for the NCs considered here, there are $H=58$ ligands adsorbed to the gold surface and that this number represents the densest packing \cite{SPV08}.
For the analysis in this appendix, we consider Lennard-Jones interaction sites as hard spheres of diameter $\sigma$.

For simplicity, we assume that the $\text{SH}$-surface is a large sphere on which the $\text{SH}$-groups are regularly distributed.
That is, one would propose that the $\text{SH}$-surface is tiled by hexagons with a $\text{SH}$-group in the center.
Then one only needs to study three $\text{SH}$-groups, and
the only unknown parameter is the distance $\varepsilon$ between the boundaries of two neighboring $\text{SH}$-spheres (cf.~Fig.~\ref{figA1}).

Assuming that the hexagonal tiling covers all the $\text{SH}$-groups distributed around the gold core at mean distance $r_\text{S}$, 
\begin{align}
4\pi r_\text{S}^2=H A\,,
\end{align}
where
$A=\frac{3\sqrt{3}}{2}\left(\frac{\sigma_\text{SH}+\varepsilon}{2}\right)^2$
represents
the hexagon area surrounding one $\text{SH}$-group, the distance between the boundaries can be estimated by
\begin{align}
\varepsilon=\sqrt{\frac{32\,\pi}{3\sqrt{3}}}\,\frac{r_\text{S}}{\sqrt{H}}-\sigma_\text{SH}=.104\,\text{nm}\,.
\end{align}

Regarding depletion, we propose that the first $\text{CH}_2$-groups (following $\text{SH}$) are the parts of the ligands that undergo depletion for small NC-distances.
Over a triangular region, as illustrated in Figure~\ref{figA1},
the hard-sphere diameters ($\sigma_{\text{CH}_2}=.396\,\text{nm}$ and $\sigma_\text{SH}=.445\,\text{nm}$ \cite{SPV08}) determine the contact distances between the first $\text{CH}_2$-group of NC\,1 (cyan) and three neighboring $\text{SH}$-groups of NC\,2 (red).
These contact distances (i.e., before depletion starts) are calculated by
\begin{align}
z(x,y)=\sqrt{\left(\frac{\sigma_\text{SH}+\sigma_{\text{CH}_2}}{2}\right)^2-x^2-(\Delta y-y)^2}\,.
\end{align}
This formula only applies to a particular part of the equilateral triangle.
Due to symmetry, it is sufficient to average over one sixth of the surface
(e.g., the triangle given by the points $(0,\Delta y),(0,0),(\Delta x,0)$ with $\Delta y=(\sigma_\text{S}+\varepsilon)/2$ and $\Delta x=\Delta y/\sqrt{3}$),
\begin{align}
\delta_\text{C)(S}=\frac{2}{\Delta x\,\Delta y}\,\int_0^{\Delta x}\int_0^{\psi(x)}z(x,y)\,dy\,dx-\delta_\text{C}-\delta_\text{S}=.300\,\text{nm}
\end{align}
(where
$\psi(x)=\Delta y-\frac{\Delta y}{\Delta x}\,x$
defines the hypotenuse of that triangle; illustrated by dots in Fig.~\ref{figA1}A).
Note that the center of the three $\text{SH}$-spheres is slightly below their mean radial distance from the gold core; the offset is given by $\delta_\text{S}=r_\text{S}-\sqrt{r_\text{S}^2-\Delta x^2-\Delta y^2}=.054\,\text{nm}$.
A similar correction is included for the $\text{CH}_2$-sphere,
$\delta_\text{C}=r_\text{C}-\frac{2}{\Delta x\,\Delta y}\,\int_0^{\Delta x}\int_0^{\psi(x)}r(x,y)\,dy\,dx=.011\,\text{nm}$,
where locally $r(x,y)=\sqrt{r_\text{C}^2-(\Delta x-x)^2-y^2}$ defines the projection of $r_\text{C}$ onto the core-to-core direction.

\section*{Acknowledgements}

Support from DFG through Collaborative Research Center 716, Project A.7 is gratefully acknowledged.
Molecular simulations were performed on the bwUniCluster.

\end{document}